\documentclass[twocolumn,aps,amsmath,showpacs]{revtex4}
\usepackage{graphicx}
\newcommand{\be}{\begin{equation}}
\newcommand{\ee}{\end{equation}}

\newcommand{\bea}{\begin{eqnarray}}
\newcommand{\eea}{\end{eqnarray}}
\newcommand{\bd}{\begin{displaymath}}
\newcommand{\ed}{\end{displaymath}}
\newcommand{\bi}{\begin{itemize}}
\newcommand{\ei}{\end{itemize}}
\newcommand{\bc}{\begin{center}}
\newcommand{\ec}{\end{center}}
\newcommand{\bfl}{\begin{flushleft}}
\newcommand{\efl}{\end{flushleft}}
\newcommand{\bfr}{\begin{flushright}}
\newcommand{\efr}{\end{flushright}}
\newcommand{\f}{\frac}

\def\6{\partial}  
  \def\ve{\varepsilon}

\def\={\!\!\!&=&\!\!\!}
\def\+{\!\!\!&&\!\!\!+~}
\def\-{\!\!\!&&\!\!\!-~}

\begin{document}
\author{S. T. Rodriguez$^1$, I. Grosu$^2$, M. Crisan$^2$, and I. \c{T}ifrea$^{1}$}

\affiliation{$^1$Department of Physics, California State University, Fullerton, CA 92834, USA}
\affiliation{$^2$Department of Physics, ``Babe\c{s}-Bolyai" University, 40084 Cluj-Napoca, Romania}

\date{\today}
\title{Thermoelectric transport properties in graphene connected molecular junctions}


\begin{abstract}
We study the electronic contribution to the main thermoelectric properties of a molecular junction consisting of a single quantum dot coupled to graphene external leads. The system electrical conductivity (G), Seebeck coefficient ($S$), and the thermal conductivity ($\kappa$), are numerically calculated based on a Green's function formalism that includes contributions up to the Hartree-Fock level. We consider the system leads to be made either of  pure or gapped-graphene. To describe the free electrons in the gapped-graphene electrodes we used two possible scenarios, the massive gap scenario, and the massless gap scenario, respectively. In all cases, the Fano effect is responsible for a strong violation of the Wiedemann-Franz law and we found a substantial increase of the system figure of merit $ZT$ due to a drastic reduction of the system thermal coefficient.  In the case of gapped-graphene electrodes, the system figure of merit presents a maximum at an optimal value of the energy gap of the order of $\Delta/D\sim$ 0.002 (massive gap scenario) and $\Delta/D\sim$ 0.0026 (massless gap scenario). Additionally, for all cases, the system figure of merit is temperature dependent. 

\end{abstract}
\maketitle

\section{Introduction}

The interest in the thermoelectric properties of low dimensional systems has recently picked up after Hicks and Dresselhaus suggested that systems with reduced dimensionality can provide the ideal candidates for efficient thermoelectric applications \cite{hicks1,hicks2}. Multi-layer quantum well superlatices, carbon-based two dimensional systems, one dimensional quantum wires, and quantum dot systems,  all have been intensively investigated both theoretically and experimentally in connection with their electronic and thermal transport properties \cite{wang,diventra}. In particular, molecular junctions consisting of quantum dots sandwiched between conducting electrodes provide various ways to control thermoelectric properties \cite{zimbovskaya}. One can control the type of material used for the electrodes, the coupling between the electrodes and the quantum dots, the Coulomb interaction between localized electrons in the quantum dots, the characteristic energy levels in the component quantum dots, and even the possible quantum interference effects in the system. Additionally, phonons in the component quantum dots play an important role in the system transport properties.

The efficiency of the thermoelectric transport is measured by the figure of merit, $ZT=G S^2 T/\kappa$, and it involves the system electrical conductivity ($G$), Seebeck coefficient ($S$), thermal conductivity ($\kappa=\kappa_e+\kappa_p$; $\kappa_e$ and $\kappa_p$ are electronic and phononic contributions), and the temperature ($T$). A large value for $ZT$ implies greater efficiency of heat-to-energy conversion. Unfortunately, standard metals obey the Wiedemann-Franz law ($GT/\kappa_e=const.$) and  an increase in their electrical conductivity leads to a reduction of their Seebeck coefficient, implying that their figure of merit is relatively small ($ZT<1$) \cite{ashcroft}. Hicks and Dresselhaus showed that $ZT$ is enhanced in systems with reduced dimensionality, i.e., $ZT\simeq 7$ in a quantum-well structure \cite{hicks1} and $ZT\simeq 14$ in a one-dimensional nanowire \cite{hicks2}. Quantum dot molecular junctions are another class of low dimensional systems with enhanced thermoelectric properties. For example, in a double quantum dot system $ZT\simeq 300$ in the vicinity of a Fano resonance when we neglect the onsite Coulomb interaction within the system component quantum dots \cite{trocha}. Such a high figure of merit corresponds to an ideal configuration of the molecular junction, however, even for more realistic situations $ZT>1$ \cite{trocha,monteros}.

To estimate the figure of merit in a molecular junction one has to calculate the electron system response to thermal and electrical driving forces. In general, when subject to a temperature gradient, $\Delta T$, a current will be induced through the system (Seebeck effect). This effect can be evaluated in the linear response regime when $\Delta T\ll T$, $T$ being the system temperature (the system leads are assumed to be at $T-\Delta T/2$ and $T+\Delta T/2$). Theoretically, the investigation of the thermoelectric transport requires the evaluation of the system electronic transmission function \cite{kim,datta}. In the case of molecular junctions, the electronic transmission function is given by the electron Green's function corresponding to the system main quantum dot, the one connected to the external leads. Various levels of approximation can be made to account for the electron correlation function in the system main dot \cite{hewson1,tifrea}. Based on the approximation level, the estimated Green's function can account for the Fano effect (lower level approximation) \cite{fano} and the Kondo effect \cite{hewson2} (higher level approximation). Additionally, other methods can be used to investigate thermoelectric effects in molecular junctions \cite{beenakker,yang}.

Here, we will focus on a simple molecular junction consisting of a single quantum dot connected to external leads. The particularity of our system is that the connecting leads are made of graphene. This molecular junction will allow us to  investigate various ways to control the system thermoelectric properties based on the coupling between the external leads and the quantum dot. Graphene, the material we considered for the molecular junction leads, is a two-dimensional carbon based material that was intensively researched due to its unusual physical properties \cite{castroneto}. In particular, graphene thermoelectrical properties have attracted a lot of attention recently  due their potential for heat-to-energy conversion \cite{balandin,hossain,chien,duan}. For the connecting leads, we 
will consider both the case of pure graphene and gapped-graphene. On the other hand, for the molecular junction component dot, we will consider a simple quantum dot modeled as a single energy level. Our analysis will be carried at the lowest level of approximation (we are neglecting the onsite Coulomb interaction in the system component dot) and our main focus  will be on the effect of the leads-quantum dot coupling.

The paper is organized as follows. In the second section we present our model and we outline the main ingredients for the evaluation of the thermoelectric properties in the system. Sec. III present our numerical results for the main thermoelectric coefficients and discuss possible ways to control the system figure of merit. Finally, in the last section we summarize our work.

\section{Thermoelectric Properties}

The molecular junction consisting of a single quantum dot sandwiched between two graphene electrodes can be described by the so-called Anderson pseudogap Hamiltonian \cite{glossop,zhu}:
\begin{eqnarray}\label{hamiltonian}
H&=&\sum_{s,\sigma;\alpha}\int_{-k_c}^{k_c}dk(\varepsilon_k-\mu)c^\dagger_{sk\sigma;\alpha}c_{sk\sigma;\alpha}\nonumber\\
&+&\sum_\sigma\varepsilon_{d}d_\sigma^\dagger d_\sigma+Un_{d\uparrow}n_{d\downarrow}
\nonumber\\
&+&\tilde{V_0}\sum_{s\sigma;\alpha}\int_{-k_c}^{k_c}dk\sqrt{|k|}(c^\dagger_{sk\sigma;\alpha}d_\sigma+d^\dagger_\sigma c_{sk\sigma;\alpha})\;.
\end{eqnarray}
The first term in the Hamiltonian describes the free electrons in the graphene leads; $c^\dagger_{sk\sigma;\alpha}$ and $c_{sk\sigma;\alpha}$ are the fermionic creation and annihilation  operators for electrons with momentum $k$ and spin $\sigma$ ($s$ stands for the valley index) in the lead $\alpha$ ($\alpha\equiv$ left (L), right (R)) and $k_c$ is a momentum cutoff. $\varepsilon_k$ stands for the electron dispersion in graphene samples and $\mu$ for the chemical potential (in general, the chemical potential can be controlled via various methods, we will consider $\mu=0$). The next two terms in the Hamiltonian describe localized electrons in the system quantum dot with characteristic energy $\varepsilon_{d}$ and subject to onsite Coulomb interaction $U$ ($d^\dagger_\sigma$ and $d_\sigma$ are fermionic creation and annihilation operators for localized electrons in the quantum dot, and $n_{d\sigma}=d_{\sigma}^\dagger d_\sigma$ is the number operator for electrons with spin $\sigma$ on the energy level $\varepsilon_{d\sigma}$). Finally, the last term in the Hamiltonian describe the hybridization of the external graphene leads into the localized energy level in the quantum dot ($\tilde{V}_0=V_0\sqrt{\pi\Omega_0}/2\pi$, $V_0$ being the tunneling amplitude and $\Omega_0$ the area of the graphene unit cell).

As we already mentioned, our molecular junction external leads are graphene based. One possibility is to consider monolayer graphene samples with low-energy electron excitations  following a linear dispersion in the vicinity of the Fermi points \cite{wallace}
\begin{equation}\label{dispersion}
\varepsilon_k=\pm\hbar  v_F k \;,
\end{equation}
with Dirac fermions moving at a speed $v_F$ about 300 times smaller than the speed of light \cite{castroneto,dassarma}. The electronic properties of monolayer graphene are those of a semimetal and they are due to a band structure that exhibits a Fermi surface that is reduced to the two points. The corresponding electron density of states has a simple linear form \cite{castroneto}
\begin{equation}\label{DOSgraphene}
\rho(E)=\rho_0|E|\;,
\end{equation}
where $\rho_0=\Omega_0/2\pi v_F^2$. On the other hand, from the application point of view, systems that present a gap between the conduction and the valence bands perform better. One can open such a gap in graphene systems using geometrical confinement \cite{geim,trauzettel,nakada,brey,chen,han}. As a different approach, a gap can be opened in graphene electronic spectrum when samples are grown on top a SiC substrate \cite{zhou1,zhou2}. Theoretically, the presence of an energy gap, $\Delta$, in graphene energy spectrum has been modeled in two different ways. The first model was introduced in connection with ARPES data and it is relatively successful close to the Dirac points. In this model, the electron energy dispersion is given by
\begin{equation}\label{massive}
\varepsilon_k^{ms}=\pm\sqrt{(v_F k)^2+\Delta^2}\;.
\end{equation}
In this case the system density of states can be obtained as \cite{koshino,ludwig}:
\begin{equation}\label{DOSmassive}
\rho_{ms}(E)=\rho_0|E|\Theta(E^2-\Delta^2)\;,
\end{equation}
with $\Theta(x)$ being the standard step function. The model, known as the massive gap model, comes short in describing the electron behavior far from the Dirac points, where according to theory graphene electrons should acquire a finite mass, $E^{ms}_{\pm}(k)\simeq\pm(\Delta+k^2/2m_{eff})$ ($m_{eff}=\Delta^2/v_F^2$). This is in contrast with experimental ARPES data that show more of a linear, massless, energy dispersion \cite{bostwick1,bostwick2}. The second model, known as the massless gapped model, tries to reconcile the gapped nature of the system electronic spectrum and the massless character of the fermions in gapped graphene \cite{benfatto}. In this model, the electron energy dispersion is given by
\begin{equation}\label{massless}
\varepsilon_k^{ml}=\pm(v_F k+\Delta)\;,
\end{equation}
with a corresponding density of states
\begin{equation}\label{DOSmassless}
\rho(E)=\rho_0\left(|E|-\Delta\right)\Theta(|E|-\Delta)\;.
\end{equation}
The model is based on a phenomenological structure of the system self-energy, but it has the advantage of conserving the massless character of the graphene electrons.

In the linear response theory, the system thermoelectric coefficients can be evaluated using the general function $L_{n\sigma}$ \cite{trocha,kim}:
\begin{equation}
L_{n\sigma}=-\f{1}{h}\int d\ve (\ve-\mu)^n\f{\partial f}{\partial \ve} T_\sigma(\ve)\;,
\end{equation}
where $f(\varepsilon)$ is the Fermi-Dirac distribution function and $T_\sigma(\ve)$ is the system transmission coefficient ($h$ is the standard Planck constant).  The electrical conductivity $G$, a measure of how easily electrons flow  in the system as a result of an external electrical potential, is given by
\begin{equation}\label{conductivity}
G=e^2\sum_\sigma L_{0\sigma}\;.
\end {equation}
The Seebeck coefficient $S$ measures the induced voltage across a system subject to a temperature gradient and can be calculated as
\begin{equation}\label{seebeck}
S=-\f{1}{eT}\f{\sum_\sigma L_{1\sigma}}{\sum_\sigma L_{0\sigma}}\;.
\end{equation}
Finally, the electronic thermal conductivity $\kappa_e$ measures how the system transfers heat when subject to a temperature gradient
\begin{equation}\label{thermalconductance}
\kappa_e=\f{1}{T}\left[\sum_\sigma L_{2\sigma}-\f{\left(\sum_\sigma L_{1\sigma}\right)^2}{\sum_\sigma L_{0\sigma}}\right]\;.
\end{equation}
(Note that in our calculation for the system thermoelectric properties we only considered the electronic contribution to the system thermal conductivity and we neglect other possible effects related to phonons.) As we already mentioned, these three parameters combine into the system figure of merit $ZT=G S^2 T/\kappa$, and allow us to estimate the system total thermoelectric efficiency.

\begin{figure}[t]
\centering \scalebox{0.9}[0.9]{\includegraphics*{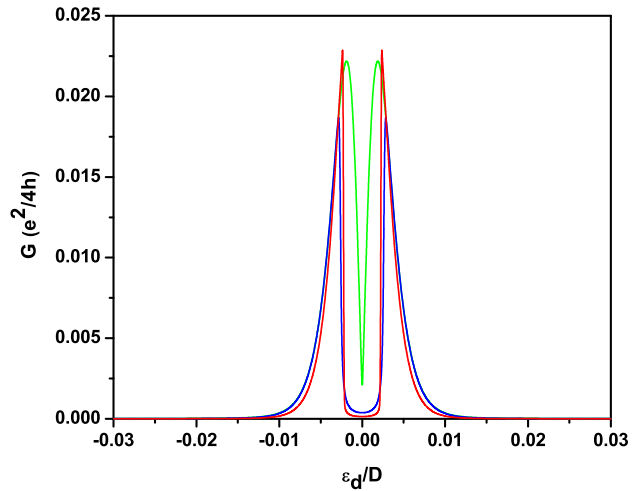}}
\caption{(Color online) The system electron conductivity ($G$) as function of the energy level in the main quantum dot:  pure graphene ($\Delta/D=0$) - green line, massive gap scenario ($\Delta/D=0.002$) - blue line, massless gap scenario ($\Delta/D=0.002$) - red line.  Other parameters in the system are $U=0$, $\Gamma_0/D=0.01$,  and $k_BT/D=0.001$.}%
\label{fig1}
\end{figure}

The transmission coefficient,  $T_\sigma(\ve)$, is related to the imaginary part of the electron-electron correlation function in the system quantum dot $T_\sigma(\varepsilon)\sim {\textrm Im} \;G^\sigma_{dd}(\varepsilon)$. The calculation of the main dot Green's function $G^\sigma_{dd}(\omega)$ can be performed using the equation of motion method \cite{hewson1}. A relatively simple calculation, that accounts for terms within the Hartree-Fock approximation, leads to
\begin{equation}\label{GreenFunction}
G^\sigma_{dd}(\omega)=\f{1-\left<n_d^{-\sigma}\right>}{\omega-\varepsilon_{d\sigma}-\Sigma^\sigma_0(\omega)}+\f{\left<n_d^{-\sigma}\right>}{\omega-\varepsilon_{d\sigma}-U-\Sigma^\sigma_0(\omega)}\;,
\end{equation}
where $\left<n_d^\sigma\right>$ represents the average occupation number for the system main quantum dot, and
\begin{equation}\label{SelfEnergy}
\Sigma^\sigma_0(\omega)=\tilde{V}_0^2\sum_\alpha\int_{-k_c}^{k_c}dk\f{|k|}{\omega-\varepsilon_k}\;,
\end{equation}
where $k_c$ is a momentum cutoff. The Green's function equation (Eq. \ref{GreenFunction}) is exact within the Hartree-Fock approximation, however, due to the fact that the average value for the occupation number can be calculated based on the Green's function itself
\begin{equation}\label{avgnumber}
\left<n_{d}^{-\sigma}\right>=-\f{1}{\pi}\int f(\omega) {\textrm Im} G^{-\sigma}_{dd} (\omega) d\omega\;,
\end{equation}
the calculation has to be performed self-consistently.

\section{Numerical results}

\begin{figure}[t]
\centering \scalebox{0.9}[0.9]{\includegraphics*{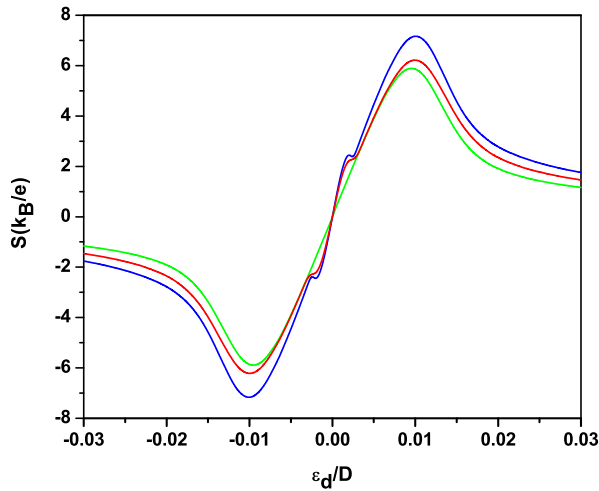}}
\caption{(Color online) The system Seebeck coefficient ($S$) as function of the energy level in the main quantum dot:  pure graphene ($\Delta/D=0$) - green line, massive gap scenario ($\Delta/D=0.002$) - blue line, massless gap scenario ($\Delta/D=0.002$) - red line.  Other parameters in the system are $U=0$, $\Gamma_0/D=0.01$,  and $k_BT/D=0.001$.}
\label{fig2}
\end{figure}

In this section we present our results for the system main thermoelectric coefficients. In general, for more reliable calculations, it is convenient to introduce dimensionless quantities and measure all energy variables in units of $D=\hbar v_Fk_c$, the cutoff energy (a reasonable value for the cutoff energy in graphene samples is $D\sim 7 eV$ \cite{isern}).  For simplicity, we will present our results for the case without onsite Coulomb interaction ($U=0$); in our conclusion part we will make additional comments regarding the role of the onsite Coulomb interaction. Unless otherwise mentioned, the system temperature is considered to be $k_BT/D=0.001$. We replace the tunneling amplitude in terms of $\Gamma_0/D=(\tilde{V}/\hbar v_F)^2$, and in our calculations we consider $\Gamma_0/D=0.01$. For comparison, we will present results for both the massive and massless gap scenarios (same value for the energy gap, $\Delta/D=0.002$) along with the pure graphene case ($\Delta/D=0$). As control parameter we will use the energy value of the localized electron level in the system quantum dot, $\varepsilon_d/D$. 

Figure \ref{fig1} presents results for the system electrical conductivity as function of the main dot localized energy for pure graphene leads ($\Delta/D=0$ - green line), and gapped graphene leads ($\Delta/D=0.002$; massive gap scenario - blue line, massless gap scenario - red line). All situations present a dip at $\varepsilon/D=0$ as a result of the Fano effect. In general,  the antiresonance due to the Fano effect happens when the main dot localized energy level aligns with the Fermi energy of the free graphene electrons ($\varepsilon_d=\varepsilon_F$). In addition, the presence of an energy gap in the molecular junction graphene leads is associated with the widening of the Fano dip in the system electrical conductivity.

\begin{figure}[t]
\centering \scalebox{0.9}[0.9]{\includegraphics*{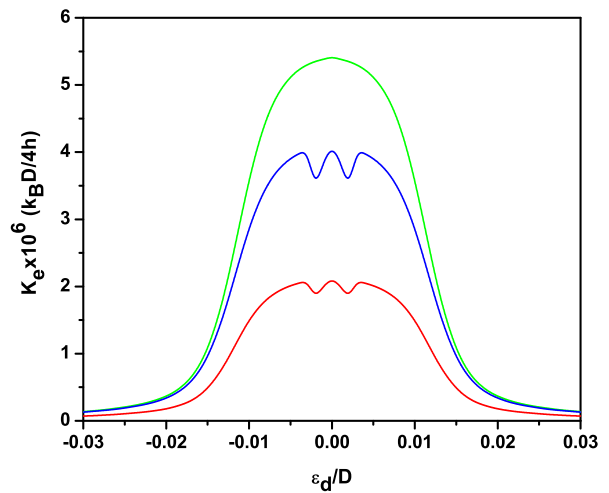}}
\caption{(Color online) The system thermal conductivity ($\kappa$) as function of the energy level in the main quantum dot:  pure graphene ($\Delta/D=0$) - green line, massive gap scenario ($\Delta/D=0.002$) - blue line, massless gap scenario ($\Delta/D=0.002$) - red line.  Other parameters in the system are $U=0$, $\Gamma_0/D=0.01$,  and $k_BT/D=0.001$.}
\label{fig3}
\end{figure}

Figure \ref{fig2} presents the system Seebeck coefficient as function of the main dot localized energy for pure graphene leads ($\Delta/D=0$ - green line), and gapped graphene leads ($\Delta/D=0.002$; massive gap scenario - blue line, massless gap scenario - red line). The shape of all curves is very similar, with small differences for the case of the gapped graphene. The signed of the Seebeck coefficient changes as the system carriers switch from electrons (positive Seebeck coefficient) to holes (negative Seebeck coefficient). When $\varepsilon_d=\varepsilon_F$, the compensation between charge currents due to electrons and holes leads to no net current and no net voltage drop in the system, therefor the Seebeck coefficient vanishes at this point.

Figure \ref{fig3} presents the system thermal conductivity $\kappa$ as function of the main dot localized energy for pure graphene leads ($\Delta/D=0$ - green line), and gapped graphene leads ($\Delta/D=0.002$; massive gap scenario - blue line, massless gap scenario - red line). Again, all three curves have a similar shape, however, the value of the system thermal conductivity is drastically reduced. Additionally, the effect of the energy gap in graphene free electron spectrum varies depending on the approximation. The highest thermal conductivity is obtained for pure graphene leads, $\Delta/D=0$. When a gap is considered, the thermal conductivity is higher in the massive gap scenario compared to the massless gap scenario. Note that in all calculations we omitted the phonon contribution to the thermal conductivity, an approximation that is valid in the low temperature limit or in the case of special engineered graphene structures \cite{sevincli}.

\begin{figure}[t]
\centering \scalebox{0.9}[0.9]{\includegraphics*{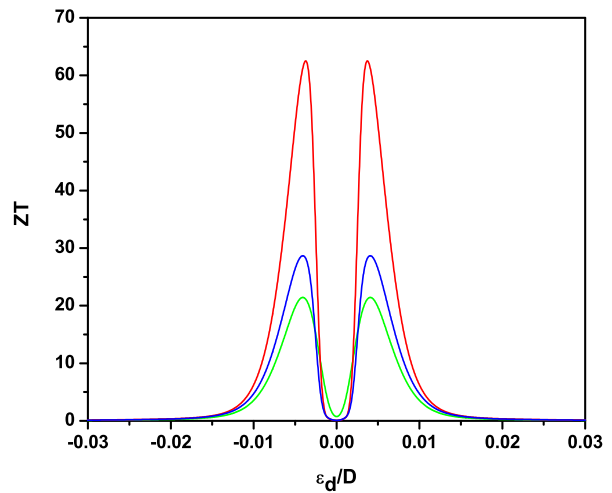}}
\caption{(Color online) The system figure of merit $ZT$ as function of the energy level in the main quantum dot:  pure graphene ($\Delta/D=0$) - green line, massive gap scenario ($\Delta/D=0.002$) - blue line, massless gap scenario ($\Delta/D=0.002$) - red line.  Other parameters in the system are $U=0$, $\Gamma_0/D=0.01$,  and $k_BT/D=0.001$}
\label{fig4}
\end{figure}

An interesting feature occurs in both the Seebeck coefficient and the thermal conductivity of the system when the value of the localized energy level $\ve_d$ is closed to the value of the energy gap in the graphene leads $\Delta$. Both coefficients present local extreme values (maxima/minima) at these points. Theoretically, this feature is related to the function $L_1$, and it is related to the step-like change in the systems density of states in the presence on an energy gap. In real systems, a step-like density of states is hard to engineer, more likely the real density of states will have a somehow smoother transition at energies close to the gap value. Accordingly, we expect that the local extreme features in the Seebeck coefficient and thermal conductivity will not be detectable experimentally.

The complexity of the system thermoelectric coefficients combines in the system figure of merit, $ZT$. In particular, the drastic reduction in the system thermal conduction will reflect into a strongly enhanced figure of merit. As one can clearly see from Figures \ref{fig1} and \ref{fig3} the molecular junction electrical and thermal conductivities behave differently, a result that it is in strong contrast  with the Wiedemann-Franz law. Figure \ref{fig4} presents the system figure of merit as function of the main dot localized energy for pure graphene leads ($\Delta/D=0$ - green line), and gapped graphene leads ($\Delta/D=0.002$; massive gap scenario - blue line, massless gap scenario - red line). In the absence of onsite Coulomb interaction the maximum values of $ZT$ are relatively high, with values of the order of $ZT_{max}\sim 20$ for pure graphene electrodes, $ZT_{max}\sim 30$ for gapped graphene electrodes within the massive gap scenario, and $ZT_{max}\sim 65$ for gapped graphene electrodes within the massless gap scenario. The effect of opening a gap in the graphene free electron energy spectrum is an enhancement of the system figure of merit due to a even stronger reduction of the system thermal conductivity. Between the two possible scenarios, the massless gap scenario seems to be more sensitive to the gap opening and implicitly the system figure of merit is higher in this case. 

\begin{figure}[t]
\centering \scalebox{0.9}[0.9]{\includegraphics*{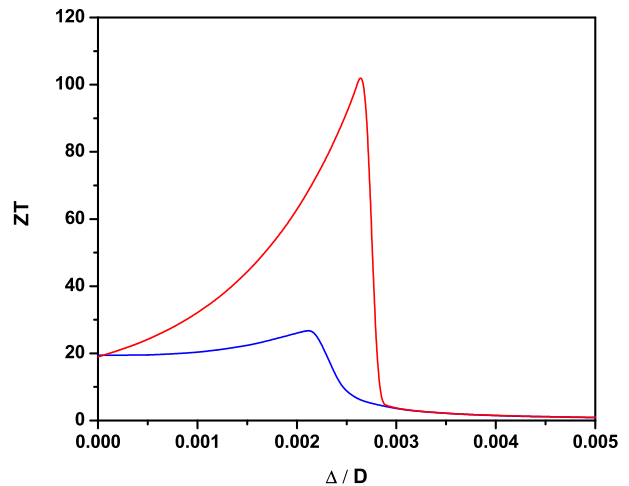}}
\caption{(Color online) The system figure of merit $ZT$ as function of the gap energy $\Delta/D$ (massive gap scenario - blue line, massless gap scenario - red line).  Other parameters in the system are $U=0$, $\Gamma_0/D=0.01$,  $\varepsilon_d/D=0.003$, and $k_BT/D=0.001$}
\label{fig5}
\end{figure}

The system thermoelectrical properties can be optimized in various ways. One choice for our molecular junction will be to engineer the energy gap in the connecting leads. Figure \ref{fig5} presents the system figure of merit $ZT$ as function of the gap energy $\Delta/D$ in the two proposed scenarios (massive gap scenario - blue line, massless gap scenario - red line). For both situations the system figure of merit presents a maximum at an optimal value of the energy gap. In the case of the massive gap scenario $ZT_{max}\sim 28$ is reached at about $\Delta/D\sim 0.002$ and for the case of the massless gap scenario $ZT_{max}\sim 105$ is reached at $\Delta/D\sim 0.0026$. As the gap increases above these values the system figure of merit drops due to a lack of thermoelectric conduction in the system.

\begin{figure}[t]
\centering \scalebox{0.9}[0.9]{\includegraphics*{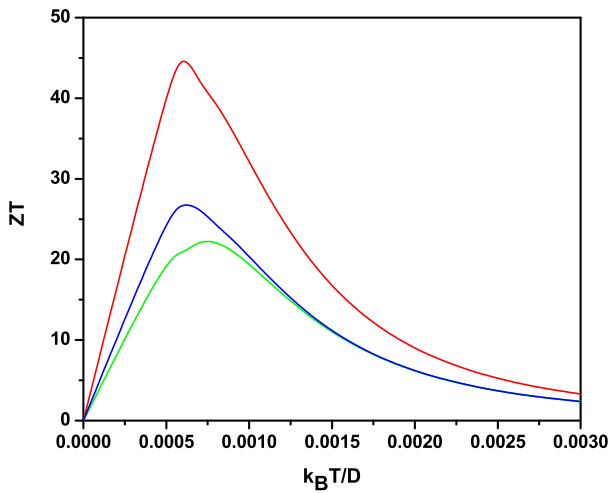}}
\caption{(Color online) The system figure of merit $ZT$ as function of temperature $k_BT/D$ (pure graphene - green line, massive gap scenario - blue line, massless gap scenario - red line).  Other parameters in the system are $U=0$, $\Gamma_0/D=0.01$,  $\varepsilon_d/D=0.003$, and $\Delta/D=0.001$.}
\label{fig6}
\end{figure}

Finally, the system figure of merit is temperature dependent. Figure \ref{fig6} presents the system figure of merit $ZT$ as function of temperature $k_BT/D$ for pure graphene leads ($\Delta/D=0$ - green line), and gapped graphene leads ($\Delta/D=0.001$; massive gap scenario - blue line, massless gap scenario - red line). For pure graphene ($\Delta/D=0$) the figure of merit presents a maximum at a value of $k_BT/D\sim 0.0007$, although for gapped-graphene the maximum is shifted to lower temperature values, $k_BT/D\sim 0.00065$. Additionally, the maximum value for  $ZT$ will be different in each situation ($ZT_{max}\sim 22$ - pure graphene leads, $ZT_{max}\sim 26$ - gapped graphene leads within the massive gap scenario, and  $ZT_{max}\sim 43$ - gapped graphene leads within the massless gap scenario). The temperature dependence of the system figure of merit is due to the different temperature dependences of the thermal coefficients that combine in the definition of $ZT$.

Onsite Coulomb interaction plays an important role in molecular junctions as it strongly influences their thermoelectric properties. As we already seen, in the absence of the Coulomb interaction ($U=0$), due to the Fano effect all the thermoelectric coefficients and implicitly the system figure of merit develop a nontrivial structure around the point $\varepsilon_d=\varepsilon_F$, i.e., when the localized level in the component quantum dot matches the Fermi energy of the free electrons in the connecting leads. In the presence of onsite Coulomb interaction ($U\neq 0$) an additional structure will develop in the systems thermoelectric coefficients around $\varepsilon_d=-U$. This structure will be very similar to the one observed in the $U=0$ case, it will consist of two peaks (this time the peaks will be asymmetric) with a sharp dip at $\varepsilon_d=-U$ due to the Fano antiresonance effect. Additionally, the onsite Coulomb interaction will have a diminishing effect on the ideal numbers we obtained for the system figure of merit. Although about an order of magnitude lower respect to the $U=0$ case, the system figure of merit will still be relevant from the practical applications point of view ($ZT>1$ even for $U\neq 0$). A similar behavior was recently reported for the case of a molecular junction consisting on a single energy quantum dot connected to external pure graphene leads  in Ref. \cite{isern}.

\section{Conclusions}

In conclusion we have analyzed the thermoelectric properties of a molecular junction consisting of one quantum dot connected to external graphene leads. Our analysis was focused on possible ways to manipulate the system properties based on the choice of graphene we are using in the connecting leads. We considered both pure graphene and gapped graphene leads for the system. Our main result is an enhancement of the system figure of merit $ZT$ when gapped graphene is used in the connecting leads. 

We considered the system main thermoelectric coefficients in the ideal limit without onsite Coulomb interaction in the component quantum dot for both pure graphene and gapped graphene leads. As a general result, the system electrical conductivity presents a dip at $\varepsilon_d=\varepsilon_F$, and at the same point, due to the compensation between electron and hole currents, the system Seebeck coefficient vanishes. Around $\varepsilon_d=\varepsilon_F$, both the system electrical conductivity and Seebeck coefficient are enhanced, although the values of these coefficients are not drastically different between the pure graphene and gapped graphene cases. On the other hand, significant differences between these cases can be observed in the system thermal conductivity. When graphene or gapped graphene leads are used instead of regular metallic leads, the system thermal coefficient is reduced by few orders of magnitude, with a significant stronger reduction for the case with gapped graphene connecting leads. A similar behavior of the system thermal coefficient was reported for graphene \cite{jiang} and for ZrS$_2$ \cite{Lv}. 

The system figure of merit is the measure of heat-to-energy conversion efficiency. The values we obtained in our calculations for both pure graphene and gapped graphene cases are consistently larger than the one previously reported in the literature, even if we considered the onsite Coulomb interaction, $ZT>1$. The most important result of our calculations is the enhancement of the system figure of merit when an energy gap is opened in the leads free electron energy spectrum.  $ZT$ is enhanced from about $ZT_{max}\sim$ 20 for pure graphene electrodes, to $ZT_{max}\sim$ 30 for gapped graphene electrodes within the massive gap scenario, and $ZT_{max}\sim$ 65 for gapped graphene electrodes within the massless gap scenario (see Fig. \ref{fig4}). The system figure of merit has a maximum at an optimal value of the energy gap, $\Delta/D~\sim0.002$ (massive gap scenario) and $\Delta/D\sim 0.0026$ (massless gap scenario), respectively. Within our approximation, the system figure of merit is temperature dependent with different maximum values depending on the theoretical scenario we considered (see Fig. \ref{fig6}).

\begin{acknowledgments}
IT would like to acknowledge financial support from the Incentive Grant program at CSUF. 
\end{acknowledgments}

\end{document}